\documentclass[12pt,english]{article}
\usepackage[latin1]{inputenc}
\usepackage{amssymb}

\makeatletter
\newcommand{\lyxaddress}[1]{
\par {\raggedright #1
\vspace{1.4em}
\noindent\par}
}

\usepackage{slashed}

\usepackage{setspace}



\makeatother

\usepackage{babel}

\begin{document}

\title{Causality in 1+1
 Dimensional Yukawa Model-II}

\author{ Asrarul Haque and Satish D. Joglekar}

\maketitle

\lyxaddress{Department of Physics, I.I.T. Kanpur, Kanpur 208016 (INDIA)}
\begin{abstract}
We discuss the limits $g\to large$, $M\to large$ with
$\frac{g^3}{M}=const.$  of the $1+1$ dimensional Yukawa model. We
take into account conclusion of the results on bound states of the
Yukawa Model in this limit (obtained in \cite{HJ09}). We find that
model reduces to an effective nonlocal $\phi^3$ theory in this
limit. We observe causality violation in this limit. We discuss the
result.
\end{abstract}
\section{Introduction}
 The (local) Standard model (SM) has, in particular, been looked
upon \cite{W} as an effective field theory, an approximation to the
actual field theory. The underlying theory could, for example, be a
composite model and the observed SM particles then may be composites
of the underlying constituents \cite{CM}. Thus, then SM particles
will generally have a finite size and are hence likely to exhibit
non-local interactions. The local SM then is an approximation; as
the effects of non-locality are such as to be normally ignorable at
the present energies. Causality violations are likely to be
associated with non-local interactions \cite{JJ04}. At least, there
are likely to be quantum violations of causality \cite{KW92,JJ04}.
Now, the effects of non-locality can possibly become visible at LHC.
Simple model calculations show that a causality violation effects
could be observed around energy scale $\leq\Lambda$, the mass scale
in theory such as the scale of compositeness \cite{JJ04,J07,HJ08}.\\
In this work, we would like to construct a simple model that
embodies this hypothesis and study the phenomenon. To this end, we
study a simple 1+1 dimensional Yukawa theory; in a certain limit of
its parameters. We show that there is \emph{ a limit of parameters }
of the theory so that there is a simple non-local effective field
theory. Causality violation is observed in the model.\\
 To understand how this is possible, we employ the results we have
obtained earlier \cite{HJ09}. We have studied elsewhere the
bound-state formation in the model in this limit and have shown that
the effective model can be interpreted as a field theory of a bound
state. We study causality in such a model. It is suggested that a
non-trivial mechanism that leads to bound states can lead to
causality violation.\\
 Even though the discussion is confined to Yukawa model, similar
effects should be observed in realistic composite models \cite{CM}
involving gauge fields.\\
In section \ref{ym1}, the Yukawa model in 1+1 dimensions is
discussed. Section \ref{ym2} deals with the motivation as to
causality violation in the bound states. In section \ref{ym3}
condition of causality due to Bogoliubov and Shirkov is briefly
reviewed. In section \ref{ym5} we show, by using the power counting,
that only one loop 2-point is divergent, 3-point function remains of
fixed magnitude and n-point function ($n\ge4$) will tend to vanish
in the large $M$ limit. Moreover the contribution to a given n-point
function from higher and higher loops fall off faster. Section
\ref{ym7} describes 3-point function which turns out to be an
effective nonlocal field theory in this limit. In Section \ref{ym8},
we find that the proper 4-point function vanishes in the limit $M\to
\infty$
 and 3-point function in one-loop approximation alone survives
in this limit. In Section \ref{ym9}, we calculate the commutator and
show that causality is violated in this model.
\section{The Model \label{ym1}}
 We shall consider the Yukawa model in 1+1 dimensions:
\begin{eqnarray}
\mathcal{L}=\bar{\psi}\left[i\slashed{\partial}-M+g\phi\right]\psi+\frac{1}{2}
\partial_{\mu}\phi\partial^{\mu}\phi-\frac{1}{2}m_{0}^{2}\phi^{2}\label{eq:lag}
\end{eqnarray}
We note that $\psi$ is of dimension $\frac{1}{2}$, $\phi$ is
dimensionless and $g$ and $M$ have dimension 1 each. We shall assume
that the interaction is \emph{ normal-ordered}, (so that the tadpole
diagrams are eliminated). Among the remaining diagrams, the one-loop
two-point function (i.e. self-energy) alone is divergent: See
section \ref{ym5}. We shall, to begin with, regularize it by a
cut-off $\Lambda$. We shall be interested in a particular limit of
parameters of the theory \cite{HJ09}. We shall consider the
possibility that the fermion-antifermion mass $M$ is very large and
at the same time the coupling constant $g$ is large, leading to a
large attractive Yukawa potential. We then found that \cite{HJ09}
the ground state is a non-relativistic heavy-quark-like bound state,
and it alone is stable; from among (a large number of) bound states.
We shall find that despite the large coupling, for a certain
relation between $g$ and $M$, the higher loop diagrams are all
smaller and smaller and the perturbation series is in fact
convergent. \subsection{Violation of Causality in Bound States
\label{ym2}} Suppose a scalar bound state of fermions is formed at
the CM at $\mathbf{x}$ at time $t$. Consider the commutator,
\[
[\phi(x),\phi(y)]
\]
Where $\phi(x)$ may be expressed as:
\begin{eqnarray*}
 \phi (x)& =& \int {\bar \psi (\xi )} \psi (\eta )f(\xi ,\eta )d\xi ;~~\frac{{\xi  + \eta }}{2} = x \\
  &=& \int {\bar \psi (x - w)} \psi (x + w)f(x - w,x + w)dw;~~\textup{Putting}~\xi = x -
  w
 \end{eqnarray*}
The function $f$ is related to the wavefunction of the bound state.
Now,
\begin{eqnarray*}
 \left[ {\phi (x),\phi (y)} \right] &=& \int {dwdz\left[ {\bar \psi (x - w)\psi (x + w),\bar \psi (y - z)\psi (y + z)} \right]}  \\
  &\times& f(x - w,x + w)f(y - z,y + z) \\
&=& \int {dwdz\left\{ {\bar \psi (x - w)\{ } \right.} \psi (x + w),\bar \psi (y - z)\} \psi (y + z) \\
 &-&\left. {  \bar \psi (y - z)\{ \psi (y + z),\bar \psi (x - w)\} \psi (x + w)} \right\} \\
  &\times& f(x - w,x + w)f(y - z,y + z)
\end{eqnarray*}
In the above commutator $\left[ {\phi (x),\phi (y)} \right]$ even if
$(x-y)^2$ is space-like, the fermions in the two bound states can
still be at time-like distances. So, they will contribute to the
commutator. If $(x-y)^2<0$ is varied, the region in spacetime over
which the anticommutator is nonzero will be varied along with. It is
unlikely for all values of $(x-y)^2<0$, the commutator will not be
equal to zero. [A similar argument has been briefly discussed by K.
Akama et al \cite{aka}].

\section{ Condition of Causality section \label{ym3} }
Bogoliubov and Shirkov \cite{BS} have formulated conditions for
causality. For this formulation, they introduce an $x$-dependent
coupling $g\left(x\right)$ in an intermediate stage and$ $ find:
\begin{equation}
\frac{\delta}{{\delta g(x)}}\left({\frac{{\delta S(g)}}{{\delta
g(y)}}S^{\dagger}(g)}\right)=0~~~ x\sim
y;~x_{0}<y_{0}\label{eq:Caus}\end{equation} where $x\sim y$ means
that $x$ is spacelike with respect to $y$. They expand the
$S$-operator as, \begin{eqnarray*}
 &  & S\left[g\left(x\right)\right]=\\
1 & + &
\sum_{n=1}\int\prod_{i=1}^{n}dx_{i}g\left(x_{1}\right)g\left(x_{2}\right).....g\left(x_{n}\right)\frac{S_{n}\left(x_{1},x_{2},...,x_{n}\right)}{n!}\end{eqnarray*}
and expand (\ref{eq:Caus}) in powers of $g$. Among these conditions,
the first one, in terms of the above $S_{n}'s$, is:\begin{eqnarray*}
S_{2}\left(x,y\right) & = & S_{1}\left(x\right)S_{1}\left(y\right)\qquad x_{0}>y_{0}\\
 & = & S_{1}\left(y\right)S_{1}\left(x\right)\qquad y_{0}>x_{0}\end{eqnarray*}
If for a given $x$ and $y$ with $\left(x-y\right)^{2}<0$, there are
Lorentz frames, in which either of the above conditions are
fulfilled. Assuming the covariance of the $S-$matrix, this, in
particular, implies,\begin{equation}
0=\left[S_{1}\left(x\right),S_{1}\left(y\right)\right]\qquad\left(x-y\right)^{2}<0\label{eq:c1}\end{equation}
and\begin{eqnarray} 0=H_{1}\left(x,y\right) & \equiv &
S_{2}\left(x,y\right)-T\left[S_{1}\left(x\right)S_{1}\left(y\right)\right]\qquad
x_{0}\neq y_{0}\label{eq:c2}\end{eqnarray} As
$S_{2}\left(x,y\right)$ is given by a covariant expression (say that
is obtained from a path-integral expression) which we denote as
below,\[
S_{2}\left(x,y\right)=T^{*}\left[S_{1}\left(x\right)S_{1}\left(y\right)\right]\]
 so that\begin{equation}
0=H_{1}\left(x,y\right)=T^{*}\left[S_{1}\left(x\right)S_{1}\left(y\right)\right]-T\left[S_{1}\left(x\right)S_{1}\left(y\right)\right]\label{eq:c3}\end{equation}
The above is a necessary condition for causality to hold. In other
words, if any matrix element of $H_{1}$ is non-zero, then that is
sufficient for CV.
\section{Calculations \label{ym4}}
\subsection{The Power Counting\label{ym5}}
 Consider a diagrams with $n$ external scalars, $L$ fermion
loops and $V$ vertices and $I_{F}$ internal fermion lines and
$I_{B}$ internal boson lines ($I_{F}+I_{B}\equiv I$) (There are no
scalar loops as there is no scalar interaction term.) For such a
diagram,
\begin{eqnarray}
3V & = & 2I+n\nonumber \\
L & = & I-V+1\nonumber \\
\mbox{i.e.}\;\;\;2L-I & = & 2-\frac{1}{2}(n+V)\;\;\;\mbox{and}\nonumber \\
D & = & 2L-I_{F}-2I_{B}=2L-I-I_{B}\nonumber \\
 & = & 2-\frac{1}{2}(n+V)-I_{B}\end{eqnarray}
 It also follows that, \begin{eqnarray}
V & = & 2L+n-2\nonumber \\
n+V & = & 2(L+n-1)\end{eqnarray}
 Ignoring the tadpole, the largest value that $D$ takes is zero and
is for the one loop two-point function diagram. This diagram alone
is divergent.\\
 First, consider the set of $n-$point one-loop diagrams
with $n\geq3$. For such a diagram, $V=n$ and $D=2-n$. The diagrams
are then finite and proportional to $\frac{g^{n}}{M^{n-2}}$.
Suppose, we consider the limit of large $g$ and $M$, such that
$\frac{g^{3}}{M}=constant=C$. Then, the 3-point function will remain
of fixed magnitude and the n-point function $(n\geq4)$ will tend to
vanish as $M^{2-\frac{2}{3}n}$. Moreover, for the diagrams with
higher number of loops, they will go as: \[ \frac{g^{V}}{M^{-D}}\sim
M^{\frac{1}{3}V+D}\sim
M^{-\left(\frac{L-1}{3}\right)-2\left(\frac{n-3}{3}\right)-I_{B}}\]
 Thus, the contributions to a given n-point function from higher an
higher loops fall off faster.
\subsection{Two-point Function \label{ym6}}
 We have noted that the one-loop two point function is the
only primitively divergent diagram. On calculation, we find:
\begin{eqnarray*} i\Sigma(p) & = & -g^{2}\int
d^{2}k\frac{1}{(2\pi)^{2}}\frac{Tr\left[(\slashed{k}+M)
(\slashed{k}+\slashed{p}+M)\right]}{\left(k^{2}-M^{2}\right)\left((k+p)^{2}-M^{2}\right)}\\
 & = & \frac{ig^{2}}{4\pi}\int_{0}^{1}d\alpha
 \left[\ln\left(\frac{\Lambda^{2}+M^{2}-\alpha(1-\alpha)p^{2}}
 {M^{2}-\alpha(1-\alpha)p^{2}}\right)-\frac{2\Lambda^{2}}
 {\Lambda^{2}+M^{2}-\alpha(1-\alpha)p^{2}}\right]\end{eqnarray*}
 For $p^{2}=0$, we have, \begin{eqnarray}
\Sigma(0) & = &
\frac{g^{2}}{4\pi}\left[\ln\left(\frac{\Lambda^{2}+M^{2}}
{M^{2}}\right)-\frac{2\Lambda^{2}}{\Lambda^{2}+M^{2}}\right]\end{eqnarray}
We next look at the correction when $p^{2}\neq0$. Assuming%
\footnote{ This condition requires explanation: it is given later.}%
 $p^{2}<<\Lambda^{2}<<M^{2}$, we find,\begin{eqnarray*}
-\Sigma(0) & =- & \frac{g^{2}}{4\pi}
\left[\ln\left(\frac{\Lambda^{2}+M^{2}}{M^{2}}\right)-\frac{2\Lambda^{2}}{\Lambda^{2}+M^{2}}\right]\\
 & \simeq & \frac{g^{2}}{4\pi}\left[\frac{\Lambda^{2}}{M^{2}}\right]\end{eqnarray*}
This is the first order quantum correction to $m^{2}$; and using
$g=(CM)^{1/3}$, and $\Lambda\sim M^{5/9}$ this falls of as
$M^{-1/9}.$ Also, \begin{eqnarray} \Sigma(p)-\Sigma(0) & = &
\frac{g^{2}}{4\pi}\int_{0}^{1}d
\alpha\alpha(1-\alpha)\frac{p^{2}}{M^{2}}\left[1-\frac{M^{2}}
{\Lambda^{2}+M^{2}}+\frac{2\Lambda^{2}M^{2}}{\left(\Lambda^{2}+M^{2}\right)^{2}}\right]\end{eqnarray}
 Using $g=(CM)^{1/3}$, we find that this leading contribution falls
of faster than $\frac{p^{2}}{M^{2}}M^{-2/3}$. Thus, the quantum
corrections, both to $m^{2}$ and the propagator vanish in this
limit. In this we have crucially made use of $\Lambda\sim M^{5/9}$
which has arisen from our study of the bound states in this model
(\cite{HJ09}): the basic assumptions are (i) in this limit, all
states except the ground bound state are absolutely unstable, (ii)
the propagator is saturated by this state, (iii) the wavefunction is
dominated by momenta $\leq M^{5/9}$.
\subsection{Three-point Function \label{ym7}}
It turns out that the three-point function is the only $O(1)$ vertex
in this limit. It is a non-local field theory with,
 \begin{eqnarray*}
S^{(3)} & \equiv & \int d^{2}x_{2}\mathcal{L}^{(3)}\left(x_{2}\right)\\
 & = & \int d^{2}x_{1}d^{2}x_{2}d^{2}x_{3}\phi\left(x_{1}\right)\phi\left(x_{2}\right)\phi\left(x_{3}\right)F\left(x_{1}-x_{2};x_{3}-x_{2}\right)\end{eqnarray*}
with \begin{eqnarray*}
 &  & FT\left\{ F\left(x_{1}-x_{2};x_{3}-x_{2}\right)\right\} =\widetilde{F}\left[p_{1,}p_{3}\right]\\
 & = & \frac{-g^{3}}{2\pi M}\int d\alpha d\beta\frac{M^{4}}{\left[M^{2}-\alpha\left(1-\alpha\right)p_{3}^{2}-\beta\left(1-\beta\right)p_{1}^{2}-2\alpha\beta p_{1}.p_{3}\right]^{2}}\end{eqnarray*}
 Causality is preserved only if, \begin{equation}
\left[\mathcal{L}^{(3)}\left(x_{1}\right),\mathcal{L}^{(3)}\left(x_{2}\right)\right]=0\qquad
whenever\:\left(x_{1}-x_{2}\right)^{2}<0.\label{eq:caus1}\end{equation}
We note that in the strict limit $M\rightarrow\infty$,
$\widetilde{F}\rightarrow const.$ and $S^{(3)}$ becomes a local
3-point interaction, and in this limit, (\ref{eq:caus1}) is
automatically fulfilled.
\subsection{ Four-point Function \label{ym8}}
The proper four-point function behaves like $\frac{g^{4}}{M^{2}}\sim
M^{-2/3}$. In the limit $M\rightarrow\infty$, the function vanishes.
The leading term in a Taylor expansion of
$\frac{g^{4}}{M^{2}}F\left(\frac{p_{i}}{M}\right)$ is of the order
$M^{-8/3}$.\\ Three-point function in one-loop approximation is what
survives in the limit we are interested in. We find:
\begin{eqnarray*}
 i\Gamma ^{(3)} & =&  - ig^3 \int {\frac{{d^2 k}}{{(2\pi )^2 }}}
 \frac{{Tr\left[ {(\not k + M)(\not k + \not p_1  + M)(\not k + \not p_1  +
 \not p_2  + M)} \right]}}{{(k^2  - M^2 )((k + p_1 )^2  - M^2 )((k + p_1  + p_2 )^2  - M^2 )}} \\
  &=&  - 2ig^3 \int {\frac{{d^2 k}}{{(2\pi )^2 }}} \frac{{M^3  + 3Mk^2  + 2M(p_1  + p_2 ).k +
  Mp_1^2  + Mp_1 .p_2 }}{{(k^2  - M^2 )((k + p_1 )^2  - M^2 )((k + p_1  + p_2 )^2  - M^2 )}} \\
&\approx& \left( {\frac{{-g^3 }}{2\pi M}} \right)\int\limits_0^1
{du\int\limits_0^{1 - u} {dv} } \frac{{M^4 }}{{\left[ {M^2  +
u(1-u)p_1^2  + v(1-v)p_2^2  + 2uvp_1. p_2 )} \right]^2 }}
\end{eqnarray*}
\section{ Evaluation of the Commutator: $\mathbf{\left[\mathcal{L}^{(3)}\left(x\right),\mathcal{L}^{(3)}\left(y\right)\right]}$ \label{ym9}}
 We have, \begin{eqnarray*}
&&C(x,y)\\
& \equiv & \left[\mathcal{L}^{(3)}\left(x\right),\mathcal{L}^{(3)}\left(y\right)\right]\\
 & = & [\int{d\xi_{1}d\xi_{2}\varphi(x)}\varphi(x+\xi_{1})\varphi(x+\xi_{2})F(\xi_{1},\xi_{2}),\int{d\eta_{1}d\eta_{2}\varphi(y)}\varphi(y+\eta_{1})\varphi(y+\eta_{2})F(\eta_{1},\eta_{2})]\\
 & = & \int{d\xi_{1}}d\xi_{2}d\eta_{1}d\eta_{2}F(\xi_{1},\xi_{2})F(\eta_{1},\eta_{2})\left[{\varphi(x)\varphi(x+\xi_{1})\varphi(x+\xi_{2}),\varphi(y)\varphi(y+\eta_{1})\varphi(y+\eta_{2})}\right]\end{eqnarray*}
 After expanding the commutator: \begin{eqnarray*}
C(x,y) & = & \int{\left\{ {\varphi(x)\varphi(x_{1})\varphi(y)\varphi(y_{1})\left[{\varphi(x_{2}),\varphi(y_{2})}\right]+\varphi(x)\varphi(x_{1})\varphi(y)\left[{\varphi(x_{2}),\varphi(y_{1})}\right]\varphi(y_{2})}\right.}\\
 & + & \varphi(x)\varphi(x_{1})\left[{\varphi(x_{2}),\varphi(y)}\right]\varphi(y_{1})\varphi(y_{2})+\varphi(x)\varphi(y)\varphi(y_{1})\left[{\varphi(x_{1}),\varphi(y_{2})}\right]\varphi(x_{2})\\
 & + & \varphi(x)\varphi(y)\left[{\varphi(x_{1}),\varphi(y_{1})}\right]\varphi(y_{2})\varphi(x_{2})+\varphi(x)\left[{\varphi(x_{1}),\varphi(y)}\right]\varphi(y_{1})\varphi(y_{2})\varphi(x_{2})\\
 & + & \varphi(y)\varphi(y_{1})\left[{\varphi(x),\varphi(y_{2})}\right]\varphi(x_{1})\varphi(x_{2})+\varphi(y)\left[{\varphi(x),\varphi(y_{1})}\right]\varphi(y_{2})\varphi(x_{1})\varphi(x_{2})\\
 & + & \left.{\left[{\varphi(x),\varphi(y)}\right]\varphi(y_{1})\varphi(y_{2})\varphi(x_{1})\varphi(x_{2})}\right\} F(\xi_{1},\xi_{2})F(\eta_{1},\eta_{2}){d\xi_{1}}d\xi_{2}d\eta_{1}d\eta_{2}\end{eqnarray*}
 Where $x_i=x+\xi_i$ and $y_i=y+\eta_i$. The last term reads: \begin{eqnarray*}
C_{9}(x,y)\equiv
\int{{\left[{\varphi(x),\varphi(y)}\right]\varphi(y_{1})\varphi(y_{2})\varphi(x_{1})\varphi(x_{2})}F(\xi_{1},\xi_{2})F(\eta_{1},\eta_{2}){d\xi_{1}}d\xi_{2}d\eta_{1}d\eta_{2}}\end{eqnarray*}
 $C_{9}(x,y)$ is zero for $\left[{\varphi(x),\varphi(y)}\right]$
vanishes for $x \sim y$. We are now left with the eight terms:
\begin{eqnarray*}
C(x,y) & = & \int{{d\xi_{1}}d\xi_{2}d\eta_{1}d\eta_{2}F(\xi_{1},\xi_{2})F(\eta_{1},\eta_{2})}\\
 & \times & \left\{ {\varphi(x)\varphi(x_{1})\varphi(y)\varphi(y_{1})\left[{\varphi(x_{2}),\varphi(y_{2})}\right]+\varphi(x)\varphi(x_{1})\varphi(y)\varphi(y_{2})\left[{\varphi(x_{2}),\varphi(y_{1})}\right]}\right.\\
 & + & \varphi(x)\varphi(x_{1})\varphi(y_{1})\varphi(y_{2})\left[{\varphi(x_{2}),\varphi(y)}\right]+\varphi(x)\varphi(y)\varphi(y_{1})\varphi(x_{2})\left[{\varphi(x_{1}),\varphi(y_{2})}\right]\\
 & + & \varphi(x)\varphi(y)\varphi(y_{2})\varphi(x_{2})\left[{\varphi(x_{1}),\varphi(y_{1})}\right]+\varphi(x)\left[{\varphi(x_{1}),\varphi(y)}\right]\varphi(y_{1})\varphi(y_{2})\varphi(x_{2})\\
 & + & \left.\varphi(y)\varphi(y_{1})\varphi(x_{1})\varphi(x_{2})\left[{\varphi(x),\varphi(y_{2})}\right]+\varphi(y)\varphi(y_{2})\varphi(x_{1})\varphi(x_{2})\left[{\varphi(x),\varphi(y_{1})}\right]\right\} \end{eqnarray*}
Let us consider the first three terms of $C(x,y)$ together:
\begin{eqnarray}
C_{123}(x,y) & \equiv& \int{{d\xi_{1}}d\xi_{2}d\eta_{1}d\eta_{2}F(\xi_{1},\xi_{2})F(\eta_{1},\eta_{2})}\nonumber\\
 & \times & \left\{ {\varphi(x)\varphi(x_{1})\varphi(y)\varphi(y_{1})\left[{\varphi(x_{2}),\varphi(y_{2})}\right]+\varphi(x)\varphi(x_{1})\varphi(y)\varphi(y_{2})\left[{\varphi(x_{2}),\varphi(y_{1})}\right]}\right.\nonumber\\
 & + & \left.\varphi(x)\varphi(x_{1})\varphi(y_{1})\varphi(y_{2})\left[{\varphi(x_{2}),\varphi(y)}\right]\right\} \label{eq:caus2}\end{eqnarray}
 The scalar field operators can be expanded in terms of creation and
annihilation operators. Each term in the r.h.s. of (\ref{eq:caus2})
will have sixteen independent terms comprising creation and
annihilation operators. $C_{123}(x,y)$ can be therefore expressed as
a collection of the sixteen independent terms. Non-vanishing of any
term for $x\sim y$ will signal causality violation. Let us consider
the specific term belonging to $C_{123}(x,y)$ as follows:\\
\begin{eqnarray*}
&& C'_{123}(x,y)\\
 && = \int{\frac{{dp_{1}dp_{2}dq_{1}dq_{2}dq_{3}}}{{(2\pi)^{5}\sqrt{2\omega_{p_{1}}2\omega_{p_{2}}}}}}\left[{\frac{a_{p_{1}}^{\dagger}a_{p_{2}}^{\dagger}a_{q_{1}}^{\dagger}a_{q_{2}}^{\dagger}}{{\sqrt{2\omega_{q_1}2\omega_{q_2}}2\omega_{q_{3}}}}}\right.e^{ip_{1}x+ip_{2}x_{1}+iq_{1}y+iq_{2}y_{1}}\sin(q_{3}(x-y+\xi_{2}-\eta_{2}))\\
 && +~ \frac{1}{{\sqrt{2\omega_{q_1}2\omega_{q_3}}2\omega_{q_{2}}}}a_{p_{1}}^{\dagger}a_{p_{2}}^{\dagger}a_{q_{1}}^{\dagger}a_{q_{3}}^{\dagger}e^{ip_{1}x+ip_{2}x_{1}+iq_{1}y+iq_{3}y_{2}}\sin(q_{2}(x-y+\xi_{2}-\eta_{1}))\\
 && +~ \left.{\frac{1}{{\sqrt{2\omega_{q_2}2\omega_{q_3}}2\omega_{q_{1}}}}a_{p_{1}}^{\dagger}a_{p_{2}}^{\dagger}a_{q_{2}}^{\dagger}a_{q_{3}}^{\dagger}e^{ip_{1}x+ip_{2}x_{1}+iq_{2}y_{1}+iq_{3}y_{2}}\sin(q_{1}(x-y+\xi_{2}))}\right]\\
 &&\times~(-2i) F(\xi_{1},\xi_{2})F(\eta_{1},\eta_{2})d\xi_{1}d\xi_{2}d\eta_{1}d\eta_{2}\end{eqnarray*}
 We have at $x^{0}=0$ and $y^{\mu}=0$: \begin{eqnarray*}
 &  & C'_{123}(x^{0}=0,x,y^{\mu}=0)\\
 & &=~ \int{\frac{{dp_{1}dp_{2}dq_{1}dq_{2}dq_{3}}}{{(2\pi)^{5}\sqrt{2\omega_{p_{1}}2\omega_{p_{2}}}}}}(-)\left[{\frac{a_{p_{1}}^{\dagger}a_{p_{2}}^{\dagger}a_{q_{1}}^{\dagger}a_{q_{2}}^{\dagger}}{{\sqrt{2\omega_{q_{1}}2\omega_{q_{2}}}2\omega_{q_{3}}}}}\right.\left({\rm {e^{i(p_{1}+p_{2}+q_{3})x}\tilde{F}(-p_{2},-q_{3})\tilde{F}(-q_{2},q_{3})}}\right.\\
 && -~ \left.{\rm {e^{i(p_{1}+p_{2}-q_{3})x}\tilde{F}(-p_{2},q_{3})\tilde{F}(-q_{2},-q_{3})}}\right)+\frac{1}{{\sqrt{2\omega_{q_{1}}2\omega_{q_{3}}}2\omega_{q_{2}}}}a_{p_{1}}^{\dagger}a_{p_{2}}^{\dagger}a_{q_{1}}^{\dagger}a_{q_{3}}^{\dagger}\\
 && \times~  \left({e^{i(p_{1}+p_{2}+q_{2})x}\tilde{F}(-p_{2},-q_{2})\tilde{F}(q_{2},-q_{3})-e^{i(p_{1}+p_{2}-q_{2})x}\tilde{F}(-p_{2},q_{2})\tilde{F}(-q_{2},-q_{3})}\right)\\
 &&+~\frac{a_{p_{1}}^{\dagger}a_{p_{2}}^{\dagger}a_{q_{2}}^{\dagger}a_{q_{3}}^{\dagger}}{{\sqrt{2\omega_{q_{2}}2\omega_{q_{3}}}\omega_{q_{1}}}}\left.{(e^{i(p_{1}+p_{2}+q_{1})x}F(-p_{2},-q_{1})-e^{i(p_{1}+p_{2}-q_{1})x}F(-p_{2},q_{1}))F(-q_{2},-q_{3})}\right]\\
 && = ~ \int{\frac{{dp_{1}dp_{2}dq_{1}dq_{2}dq_{3}}}{{(2\pi)^{5}\sqrt{2\omega_{p_{1}}2\omega_{p_{2}}2\omega_{q_{1}}2\omega_{q_{2}}}}}}\frac{{a_{p_{1}}^{\dagger}a_{p_{2}}^{\dagger}a_{q_{1}}^{\dagger}a_{q_{2}}^{\dagger}}}{{2\omega_{q_{3}}}}(e^{i(p_{1}+p_{2}-q_{3})x}-e^{i(p_{1}+p_{2}+q_{3})x})\\&& \times~ \tilde{F}(-p_{2},-q_{3})\tilde{F}(-q_{2},q_{3})\\
 && +~ \int{\frac{{dp_{1}dp_{2}dq_{1}dq_{2}dq_{3}}}{{(2\pi)^{5}\sqrt{2\omega_{p_{1}}2\omega_{p_{2}}2\omega_{q_{1}}2\omega_{q_{3}}}}}}\frac{{a_{p_{1}}^{\dagger}a_{p_{2}}^{\dagger}a_{q_{1}}^{\dagger}a_{q_{3}}^{\dagger}}}{{2\omega_{q_{2}}}}(e^{i(p_{1}+p_{2}-q_{2})x}-e^{i(p_{1}+p_{2}+q_{2})x})\\ &&\times~ \tilde{F}(-p_{2},-q_{2})\tilde{F}(q_{2},-q_{3})\\
 && +~  \int{\frac{{dp_{1}dp_{2}dq_{1}dq_{2}dq_{3}}}{{(2\pi)^{5}\sqrt{2\omega_{p_{1}}2\omega_{p_{2}}2\omega_{q_{2}}2\omega_{q_{3}}}}}}\frac{{a_{p_{1}}^{\dagger}a_{p_{2}}^{\dagger}a_{q_{2}}^{\dagger}a_{q_{3}}^{\dagger}}}{{2\omega_{q_{1}}}}(e^{i(p_{1}+p_{2}-q_{1})x}-e^{i(p_{1}+p_{2}+q_{1})x})\\&& \times~ \tilde{F}(-p_{2},-q_{1})\tilde{F}(-q_{2},-q_{3})\end{eqnarray*}
 Now, we can subject Taylor expansion to $C'_{123}(x^{0}=0,x,y^{0}=0,y=0)$
around $x=0$: \begin{eqnarray*}
C'_{123}(x^{0}=0,x,y^{\mu}=0)=C'_{123}(x^{0}=0,x=0,y^{\mu}=0)+x\frac{\partial}{{\partial
x}}C'_{123}(x^{0}=0,x,y^{\mu}=0)\left|{_{x=0}}\right.+....\end{eqnarray*}
 We can have now, \begin{eqnarray*}
 &  & \frac{\partial}{{\partial x}}C'_{123}(x^{0}=0,x,y^{\mu}=0)\left|{_{x=0}}\right.\\
 &  & = ~\int{\frac{{dp_{1}dp_{2}dq_{1}dq_{2}dq_{3}}}{{(2\pi)^{5}\sqrt{2\omega_{p_{1}}2\omega_{p_{2}}2\omega_{q_{1}}2\omega_{q_{2}}}}}}\frac{{a_{p_{1}}^{\dagger}a_{p_{2}}^{\dagger}a_{q_{1}}^{\dagger}a_{q_{2}}^{\dagger}}}{{2\omega_{q_{3}}}}(-2q_{3})\tilde{F}(-p_{2},-q_{3})\tilde{F}(-q_{2},q_{3})\\
 & & + ~\int{\frac{{dp_{1}dp_{2}dq_{1}dq_{2}dq_{3}}}{{(2\pi)^{5}\sqrt{2\omega_{p_{1}}2\omega_{p_{2}}2\omega_{q_{1}}2\omega_{q_{3}}}}}}\frac{{a_{p_{1}}^{\dagger}a_{p_{2}}^{\dagger}a_{q_{1}}^{\dagger}a_{q_{3}}^{\dagger}}}{{2\omega_{q_{2}}}}(-2q_{2})\tilde{F}(-p_{2},-q_{2})\tilde{F}(q_{2},-q_{3})\\
 &  &+~ \int{\frac{{dp_{1}dp_{2}dq_{1}dq_{2}dq_{3}}}{{(2\pi)^{5}\sqrt{2\omega_{p_{1}}2\omega_{p_{2}}2\omega_{q_{2}}2\omega_{q_{3}}}}}}\frac{{a_{p_{1}}^{\dagger}a_{p_{2}}^{\dagger}a_{q_{2}}^{\dagger}a_{q_{3}}^{\dagger}}}{{2\omega_{q_{1}}}}(-2q_{1})\tilde{F}(-p_{2},-q_{1})\tilde{F}(-q_{2},-q_{3})\end{eqnarray*}
 Making the following interchanges in the above expression: $(q_{2}\leftrightarrow q_{3})$
in the second term and $(q_{1}\leftrightarrow q_{3})$ in the third
term, $\frac{\partial}{{\partial
x}}C'_{123}(x^{0}=0,x,y^{\mu}=0)\left|{_{x=0}}\right.$ takes the
form: \begin{eqnarray*}
 &  & \frac{\partial}{{\partial x}}C'_{123}(x^{0}=0,x,y^{\mu}=0)\left|{_{x=0}}\right.\\
 &&=~ \int{\frac{{dp_{1}dp_{2}dq_{1}dq_{2}dq_{3}}}{{(2\pi)^{5}\sqrt{2\omega_{p_{1}}2\omega_{p_{2}}2\omega_{q_{1}}2\omega_{q_{2}}}}}}\frac{{a_{p_{1}}^{\dagger}a_{p_{2}}^{\dagger}a_{q_{1}}^{\dagger}a_{q_{2}}^{\dagger}}}{{2\omega_{q_{3}}}}(-2q_{3})(\tilde{F}(-p_{2},-q_{3})\tilde{F}(-q_{2},q_{3})\\
 & &+~ \tilde{F}(-p_{2},-q_{3})\tilde{F}(q_{3},-q_{2})+\tilde{F}(-p_{2},-q_{3})\tilde{F}(-q_{2},-q_{1}))\end{eqnarray*}
Which is nonzero. Thus the commutator
$\left[\mathcal{L}^{(3)}\left(x\right),\mathcal{L}^{(3)}\left(y\right)\right]$
does not vanish even for space-like separation and leads to the
violation of causality in this model.
\section{ Discussion and Conclusions \label{ym10}}
We have discussed the particular limit of the Yukawa field theory in
$1+1$ dimensions. We find that all n-point function except 2-point
and 3-point functions tend to vanish into this limit. Higher order
corrections to the 2-point function corrected for using bound state
model results also tends to vanish in this limit. This theory
reduces to effective nonlocal scalar field theory. Causality
violation is observed in this model.\\
The original theory is causality preserving. So we may suspect the
causality should be preserved in the effective theory. However this
problem has been investigated recently \cite{J07}, and it has been
found that composite state model can give to the causality violation
even if underlying field theory does not show any causality
violation. The source of causality violation in this model exhibits
bound states which we have taken into account while calculating at
least propagator. This gives us only with effective $\phi^3$
interaction which is nonlocal and causality violating.\\

ACKNOWLEDGEMENT\\
Part of the work was done when SDJ was "Poonam and Prabhu Goel Chair
Professor" at IIT Kanpur.

\end{document}